\documentclass[a4paper,12pt]{article}
\usepackage{graphicx}
\usepackage{color}
\usepackage{amstext}
 \usepackage[dvipdf]{epsfig}
  \usepackage{color}
  \usepackage{amssymb}

\def\d{\textsf d}
\def\no{\nonumber}
\newcommand{\ben}{\begin{eqnarray}}
\newcommand{\een}{\end{eqnarray}}
\def\be{\begin{equation}}
\def\ee#1{\label{#1}\end{equation}}
\def\lb{\label}
\def\bx{\mathbf{x}}
\def\bq{\mathbf{q}}
\def\bc{\mathbf{c}}
\def\bv{\mathbf{v}}
\def\bk{\mathbf{k}}
\def\d{\textsf{d} }
\def\bg{\mathbf{g}}
\def\a{{\alpha }}

\begin{document}
\title{Instabilities in a self-gravitating granular gas
\\Gilberto M. Kremer\footnote{kremer@fisica.ufpr.br}\\
Departamento de F\'isica, Universidade Federal do Paran\'a, Caixa Postal 19044, 81531-990 Curitiba, Brazil}

\maketitle

\begin{abstract}
A kinetic and hydrodynamic descriptions are developed in order to analyze the instabilities in a self-gravitating granular gas. In the kinetic description the Boltzmann equation is coupled with the Poisson equation, while in the hydrodynamic description the Poisson equation is coupled with the balance equations of mass density, hydrodynamic velocity and temperature for an Eulerian fluid. In the background solution for both descriptions the fluid is at rest with constant mass density and gravitational potential while the temperature depends on time through Haff's law. In the kinetic description the perturbed distribution function and gravitational potential in the Fourier space are related to time dependent small amplitudes. In the hydrodynamic description the perturbed mass density, hydrodynamic velocity and temperature in the Fourier space are functions of time dependent small amplitudes. From the analysis of the system of coupled differential equations for the amplitudes for the two descriptions  the time evolution of the density contrast -- a parameter that indicate where there are local enhancements in the matter density -- is determined. The solutions depend on two parameters, one is the mean free path of the gas particles and another Jeans' wavelength, which is a function of the gravitational constant, mass density and speed of sound of the gas.
It is shown that instabilities due to the inelastic collisions occur when the Jeans and the perturbation  wavelengths are larger than the mean free path, while Jeans' instabilities due to the gravitational field happen when  the mean free path and the perturbation wavelength are larger than Jeans' wavelength.

\end{abstract}

\section{Introduction}

Fluid instabilities in self-gravitating gases was first studied by Jeans \cite{b1} who analyzed the  system of equations of mass and momentum densities for an Eulerian fluid  coupled with the Poisson equation. From the dispersion relation for the perturbed density contrast that followed from the system of equations he showed that apart from the harmonic perturbations there existed growing and decaying modes. In the dispersion relation he identified a cutoff -- known nowadays as Jeans' wavelength -- where for wavelengths smaller  than Jeans's wavelength the perturbations perform harmonic oscillations  whereas for large wavelengths the perturbations increase or decrease with time. The time increase of the  perturbations is known as Jeans' instability and  describes gravitational collapse of self-gravitating interstellar gas clouds \cite{b2,b3,b4,b5}.

 There is a very simple model described in the literature to understand Jeans' instability: let us consider a mass density inhomogeneity inside a mass $M$ enclosed in a volume of radius $\lambda$, the inhomogeneity will grow if the gravity force per unit mass $F_g=GM/\lambda^2\propto G\rho\lambda^3/\lambda^2$ is greater than the opposed pressure force per unit of mass $F_p\propto p\lambda^2/\rho\lambda^3$, where $G$ is the gravitational constant, $p$ the pressure and $\rho$ the mass density.
By introducing the sound velocity $v_s\propto\sqrt{ p/\rho}$ and Jeans' wavelength $\lambda_J\propto v_s/\sqrt{G\rho}$ the inhomogeneity will grow if $\lambda>\lambda_J$. As an equivalent statement we may say that the time scale of the pressure exerted in a region $t_p\propto \lambda/v_s$ must be bigger than the time scale needed to start the gravitational collapse of the matter due to its own weight $t_g\propto1/\sqrt{G\rho}$, i.e., $\lambda>\lambda_J$.

Granular gases refer to the fluid behavior in rapid flows of granular materials subjected to driven forces and described by hydrodynamic equations of motion. The particle interactions for granular gases are inelastic and the energy dissipation at collisions implies a temperature decay of the gas.
For non-self-gravitating granular gases it was shown that density fluctuations will lead to the formation of clusters \cite{Gd,Mc,Ern,Brey,Br,Gar1,Gar2} which can be understand as follows: the increase of density due to a fluctuation implies into an increase of the inelastic collisions which will decrease the temperature and as a consequence, a decrease in the pressure  happens. Hence, a pressure gradient will be established from the increased density region  to the neighboring regions so that according to Fick's law, particles will flow in opposite direction to the pressure gradient implying into a clustering of gas particles in the region where the fluctuation occurred.

Jeans' instability was also studied within the framework of Boltzmann equation coupled with Poisson equation in a static universe  \cite{b3,b4,b6,b7} and  in an expanded universe \cite{b8,b9}. In the context of alternatives theories of gravity  Jeans' instability was also investigated in refs. \cite{fr1,fr2,fr3,fr4,fr5}. Recently  Jeans' instability in a static and in an expanded universe with dissipation  was analyzed in Ref. \cite{b10} by considering the hydrodynamic equations of a five-field and a thirteen-field theories.

During the preparation of this work  a paper on Jeans' instability for granular gases appeared in the arXiv \cite{b11}. In the referred work, Jeans and clustering instabilities for a viscous and heat conducting granular gas were analyzed within a five-field theory of mass, momentum and energy densities. Apart from the analysis of the instabilities, the growths of disturbances related to the shear, sound and heat modes as functions of the wave number were determined.
Although the subject of the present work is the same as the quoted paper, the analysis of that paper is based on the hydrodynamic equations, while here a kinetic description based on the Boltzmann equation is developed. Furthermore, the hydrodynamic theory developed here has a different methodology as the quoted paper although restricted to an Eulerian granular gas.

The aim of this work is to analyze the stability of self-gravitating granular gases. Here we develop two methods, one based in a kinetic description and another in a hydrodynamic description. The kinetic description is based on the coupling of the Poisson equation with the Boltzmann equation, while in the  hydrodynamic description  the balance equations of mass density, hydrodynamic velocity and temperature for an Eulerian fluid -- where only the energy dissipation is taken into account -- are coupled with the Poisson equation. The background solution in both descriptions is characterized by a constant  mass density, a vanishing  hydrodynamic velocity and  a constant gravitational potential, while the temperature obeys Haff's law. Superposed to the background solution small perturbations of time-dependent amplitudes and Fourier space modes are considered. A coupled system of differential equations for the time-dependent amplitudes is obtained, which is a function of two parameters. One is the mean free path of the gas particles and another is Jeans wavelength, which is a function of the gravitational constant, mass density and speed of sound of the gas. As in the case of Jeans instability -- which describes gravitational instability of self-gravitating gas clouds -- for large wavelengths with respect to Jeans wavelengths the amplitudes growth exponentially implying granular gas instabilities. However, for small wavelengths with respect to Jeans wavelengths time oscillations of the amplitudes follow. It is shown that for granular gases these two behaviors depend also on the ratio of the mean free path and Jeans wavelength.

The paper is organized as follows: the kinetic and hydrodynamic descriptions are developed in Sections 2 and 3, respectively. Final remarks are given in Section 4 and the  main conclusions of the paper are summarized in Section 5.

\section{Kinetic description}
\subsection{Boltzmann equation}

The kinetic description of a granular gas is based on the Boltzmann equation which refers to the space-time evolution of the one-particle distribution function $f(\bx,\bv,t)$ in the phase space spanned by the space and velocity coordinates $(\bx,\bv)$ of the molecules. The molecules have mass $m$ and diameter $\d$ and the encounters between the molecules are inelastic so that the momentum is conserved at collision but not the energy.

If $(\bv,\bv_1)$ denote the pre-collisional velocities and $(\bv',\bv_1')$  the post-collisional velocities of two molecules at collision, the inelastic encounters are characterized by the relationship
 $(\bg^\prime\cdot\bk)=-\a (\bg\cdot\bk)$, which relates the pre-relative velocity $\bg=\bv_1-\bv$ and the post-collisional velocity $\bg^\prime=\bv_1^\prime-\bv^\prime$ at collision. The parameter $0\leq\a\leq1$ is the normal restitution coefficient and $\bk$ the unit vector directed along the line which joins the molecules centers and pointing from center of the molecule labeled by the index 1 to the center of the molecule without label. In the inelastic collisions the component of the velocity perpendicular to the collision vector $\bk$ does not change so that  $\bk\times\bg^\prime=\bk\times\bg$.

 The momentum conservation law $m\bv+m\bv_1=m\bv'+m\bv_1'$  implies the following relationships between the post- and pre-collisional velocities
 \ben\lb{k1}
 &&\bv^\prime=\bv+\frac{1+\a}2(\bg\cdot\bk)\bk,\quad
 \bv_1^\prime=\bv_1-\frac{1+\a}2(\bg\cdot\bk)\bk,
 \\\lb{k2}
 &&\bg^\prime=\bg-(1+\a)(\bg\cdot\bk)\bk,
 \een
 while the variation of the kinetic energy in terms of the pre- and post-collisional velocities becomes
 \be
 {m\over2}v^{\prime2}+{m\over2}v_1^{\prime2}-{m\over2}v^{2}-{m\over2}v_1^{2}
 ={m\over4}(\a^2-1)(\bg\cdot\bk)^2.
 \ee{k3}
In the case of elastic collisions $\a=1$ and one recovers the kinetic energy conservation law.

In a restitution collision the equations that relate the  post-collisional velocities of two molecules  $(\bv, \bv_1)$ with the pre-collisional velocities denoted by $(\bv^\ast, \bv_1^\ast)$ are
 \be
 \bv=\bv^\ast+\frac{1+\a}2(\bg^\ast\cdot\bk^\ast)\bk^\ast,\qquad
 \bv_1=\bv_1^\ast-\frac{1+\a}2(\bg^\ast\cdot\bk^\ast)\bk^\ast,
 \ee{k4}
where $\bk^\ast=-\bk$ and $(\bg\cdot\bk)=-\a(\bg^\ast\cdot\bk)$.

The modulus of the Jacobian of the transformation  $d{\bf c}_1^\ast\,d{\bf
c^\ast}=\vert J\vert d{\bf c}_1\,d{\bf c}$ is given by $\vert J\vert={1/\a}$ so that
$(\bg^\ast\cdot\bk^\ast)\,d\bc^\ast\,d{\bf c}_1^\ast={1\over\a^2}(\bg\cdot\bk)\,d\bc\,d{\bf
 c}_1$. Hence the Boltzmann equation in the presence of a gravitational potential $\phi$ reads \cite{Br,Gar2,Kr}
\ben\lb{k5}
{\partial_t f}+{\bv}\cdot\nabla f-\nabla\phi\cdot\partial_{\bv}f
=\int\left(\frac1{\a^2}f_1^*f^*-f_1f\right)\d^2(\bg\cdot\bk)d\bk d\bv_1.
\een
Here $f_1^*=f(\bx,\bv_1^*,t)$ and so on.

The Boltzmann equation in the presence of a gravitational potential is coupled with the Poisson equation for the Newtonian gravitational potential $\phi$, namely
\ben\lb{k6}
\nabla^2\phi=4\pi G\rho=4\pi G\int mfd\bv,
\een
where $G$ is the gravitational constant and $\rho$ the mass density of the granular gas .

From the Boltzmann equation we can obtain a transfer equation for an arbitrary function of the molecular velocities $\psi(\bv)$. Indeed from the multiplication of (\ref{k5}) by $\psi(\bv)$  and integration of the resulting equation over all values of the velocity $\bv$ we get
\ben\no
&&\int\psi(\bv)\left[{\partial_t f}+{\bv}\cdot\nabla f-\nabla\phi\cdot\partial_{\bv}f\right]d\bv
\\\lb{k7}
&&\qquad=\frac12\int\left[\psi(\bv')+\psi(\bv_1')-\psi(\bv)-\psi(\bv_1)\right]f_1f\d^2(\bg\cdot\bk)d\bk d\bv_1d\bv,
\een
where the symmetry properties of the collision operator were used to write the right-hand side of the above equation.

\subsection{Jeans instability}
We consider a granular gas which initially is at rest with a constant mass density $\rho_0$ and a time dependent temperature $T(t)$ subjected to a constant gravitational potential  $\phi_0$. In this case the distribution function is characterized by the Maxwellian
\ben\lb{k8}
f_0(\bv,t)=\frac{\rho_0}m\left(\frac{m}{2\pi k_BT(t)}\right)^\frac32\exp\left[-\frac{mv^2}{2k_BT(t)}\right],
\een
where $k_B$ is the Boltzmann constant.

Insertion of the Maxwellian distribution function (\ref{k8}) into the transfer equation (\ref{k7}) leads to
\be
\int\psi(\bv)\partial_t f_0d\bv=\frac12\int\!\big[\psi(\bv')+\psi(\bv_1')-\psi(\bv)-\psi(\bv_1)\big]f^1_0f_0\d^2(\bg\cdot\bk)d\bk d\bv_1d\bv.
\ee{k9a}

For the values of $\psi(\bv)$ equal to the molecular  mass $m$ and momentum $m\bv$ the transfer equation (\ref{k9a}) is identically zero, but for the molecular energy $mv^2/2$ we get after the integration of the resulting equation Haff's law
\ben\lb{k9b}
\frac{d T}{d t}=-\zeta T,\qquad \zeta=\frac43(1-\a^2)\d^2\frac{\rho_0}m\sqrt{\frac{\pi k_BT}m}.
\een
The coefficient $\zeta$ is due to the energy dissipation of the colliding gas molecules and it is known as the cooling rate. Haff's law represents the homogeneous cooling of the granular gas.

We note that the condition of a constant gravitational potential implies that $\nabla\phi_0=0$. This condition may follow  from symmetry considerations, because in a homogeneous  system  there is no preference in the direction of the gravitational potential gradient.
Nevertheless, the condition $\nabla\phi_0=0$ does not satisfy the Poisson equation (\ref{k6}), because its right-hand side is proportional to the mass density. In order to overcome this inconsistency we shall use the well-known "Jeans swindle" (see e.g. \cite{b3,b4}), which considers that the Poisson equation is valid only  for the perturbed distribution function and perturbed gravitational potential. It is interesting to note that the "Jeans swindle" is not necessary when one considers an expanded universe described by the Friedmann-Lama\^itre-Robertson-Walker metric (see e.g. \cite{b8,b9,b10}).

Superposed to the Maxwellian distribution and constant gravitational potential we introduce small perturbations of the distribution function $h(\bx,\bv,t)$ and gravitational potential $\phi_1(\bx,t)$, namely
\ben\lb{k10a}
f(\bx,\bv,t)=f_0(\bv,t)\left[1+h(\bx,\bv,t)\right],
\qquad
\phi(\bx,t)=\phi_0+\phi_1(\bx,t).
\een

The  transfer equation for the perturbations of the distribution function and gravitational potential follows from the insertion of (\ref{k10a})  into (\ref{k7}), yielding
\ben\no
\int\psi(\bv)\left[h\partial_t f_0+f_0\partial_th+f_0\bv\cdot\nabla h-\nabla\phi_1\cdot \partial_{\bv}f_0\right]d\bv\quad
\\\lb{k11}
=\frac12\int\big[\psi(\bv')+\psi(\bv_1')-\psi(\bv)-\psi(\bv_1)\big] f^1_0f_0\big[h+h_1\big]\d^2(\bg\cdot\bk)d\bk d\bv_1d\bv.
\een
Above all non-linear terms were neglected.

The perturbed Poisson equation which follows from (\ref{k6}) reads
\ben\lb{k12}
\nabla^2\phi_1=4\pi G\int mhd\bv,
\een

We suppose that the perturbations are represented by plane waves with wavenumber vector $\bq$ and time-dependent small amplitudes. Furthermore, the amplitudes of the perturbed distribution function are linear combination of the molecular mass, momentum and energy \cite{b7,b8,b10}:
\ben\lb{k12a}
&&h(\bx,\bv,t)=\big[A(t)+{\bf B}(t)\cdot\bv+D(t)v^2\big]\exp[i\bq\cdot\bx],\;
\\\lb{k12b}
&&\phi_1(\bx,t)=\phi_1(t)\exp\left[i\bq\cdot\bx\right].
\een
Here $A(t), {\bf B}(t), D(t)$ and $\phi_1(t)$ are the small time-dependent amplitudes.

If we choose $\psi(\bv)$ equal to $m, m\bv$ and $mv^2/2$ into (\ref{k11}), use the representations (\ref{k12a}), (\ref{k12b})  and integrate the resulting equations we get respectively the following system of differential equations for the amplitudes $A(t), {\bf B}(t)$ and $D(t)$:
\ben\lb{k13a}
&&\frac{d A}{d t}+3\frac{k_B}m\frac{d (DT)}{d t}+i\frac{k_BT}m\bq\cdot{\bf B}=0,
\\\lb{k13b}
&&\frac{k_B}m\frac{d({\bf B}T)}{dt}+i\frac{k_BT}m\bq\left(A+5\frac{k_B}m DT\right)+i\bq \phi_1=0,
\\\no
&&\frac{d(AT)}{dt}+5\frac{k_B}m\frac{d(DT^2)}{dt}+i\frac{5k_B}{3m}T^2\bq\cdot{\bf B}
\\\lb{k13c}
&&\qquad=-\frac43(1-\a^2)\d^2\frac{\rho_0}m\left(\frac{\pi k_B}m\right)^\frac12T^\frac32\left(2A+9\frac{k_B}m DT\right).
\een

Furthermore, after  integration the perturbed Poisson equation (\ref{k12}) with the representations (\ref{k12a}) and (\ref{k12b}) becomes
\ben\lb{k14}
-q^2\phi_1=4\pi G\rho_0\left(A+3\frac{k_B}m DT\right).
\een

The density contrast is a parameter which indicate where there are local enhancements in the matter density. Here the density contrast is given in terms of the amplitudes of the perturbed distribution function, namely $\delta_\rho=(\rho-\rho_0)/\rho_0=\left(A+3\frac{k_B}m DT\right)$.

The derivation of (\ref{k13a}) with respect to time and the eliminations of the amplitude $D(t)$, by the the use of the definition of the density contrast, and the amplitudes ${\bf B}(t)$ and $\phi_1(t)$ by the use of (\ref{k13b}) and (\ref{k14}), respectively,  we get the following differential equation for the density contrast
\ben\lb{k15a}
\frac{d^2\delta_\rho}{dt^2}+\frac{5k_B}{3m}\left(T\delta_\rho-\frac25AT\right)q^2-4\pi G\rho_0\delta_\rho=0.
\een

Likewise the elimination of $D(t)$ and ${\bf B}(t)$ from (\ref{k13c}) leads to a differential equation for the product $A(t)T(t)$:
\be
\frac{dAT}{dt}-\frac83(1-\a^2)\d^2\frac{\rho_0}m\left(\frac{\pi k_B}m\right)^\frac12T^\frac12\left(T\delta_\rho-\frac34AT\right)=0.
\ee{k15b}

For the determination of the time evolution of the density contrast we have a system of differential equations consisted of (\ref{k9b}), (\ref{k15a}) and (\ref{k15b}).
In order to solve this system of differential equations we introduce a dimensionless temperature $\widetilde T=T/T_0$ and a dimensionless time $\tau=t/t_0$ where $T_0$ is a constant reference  temperature  and $t_0=(m/ 4\rho_0\d^2)\sqrt{m/\pi k_BT_0}$ a molecular mean free time. Hence,  Haff's law (\ref{k9a}) in terms of the dimensionless quantities becomes ${d\widetilde T}/{d\tau}+\widetilde T^\frac32(1-\a^2)/3=0$ whose solution for the initial condition $\widetilde T(0)=1$ (say) is
\ben\lb{k16}
\widetilde T(\tau)=\frac1{\left[1+(1-\a^2)\tau/6\right]^2}.
\een

If we introduce a new dimensionless time  $\tau_*= t/t_g=\sqrt{4\pi G\rho_0}\,t$ -- where $t_g$ denotes the time to start the gravitational collapse -- the dimensionless equations that follow from (\ref{k15a}) and (\ref{k15b}) read
\ben\lb{k17a}
\frac{d^2\delta_\rho}{d\tau_*^2}+\left(\widetilde T\delta_\rho-\frac25A_*\right)\frac{q^2}{q_J^2}-\delta_\rho=0,
\\\lb{k17b}
\frac{dA_*}{d\tau_*}-\sqrt{\frac4{15}}(1-\a^2)\frac{\lambda_J}{l_0}\widetilde T^\frac12\left(\widetilde T\delta_\rho-\frac34A_*\right)=0.
\een
Here $l_0=m/4\sqrt\pi\rho_0 \d^2$ denotes the molecular mean free path, $q_J=\sqrt{4\pi G\rho_0}/v_s$   Jeans' wavenumber, $v_s=\sqrt{(5kT_0/3m)}$ the gas sound speed and $\lambda_J=2\pi/q_J$ Jeans' wavelength.

 Equations (\ref{k17a}) and (\ref{k17b}) constitute a system of differential equations for the determination of $\delta_\rho$ and  $A_*=A\widetilde T$ as function of the dimensionless time $\tau_*$.

Let us analyze first the case of  elastic collisions  where $\alpha=1$. In this case we have from (\ref{k16}) that $\widetilde T=1$ for the initial condition that $\widetilde T(0)=1$. Furthermore,  (\ref{k17b}) implies that $A_*$ does not depend on the dimensionless time $\tau_*$, i.e. $A_*= C$= constant. In this case the evolution equation for the density contrast (\ref{k17a}) becomes
\ben\lb{k18a}
\frac{d^2\delta_\rho}{d\tau_*^2}+\left(\frac{q^2}{q_J^2}-1\right)\delta_\rho-\frac{2C}5\frac{q^2}{q_J^2}=0.
\een

The solution of the differential equation (\ref{k18a}) is given by
\be
\delta_\rho(\tau_*)=\frac{2Cq^2/q_J^2}{5(q^2/q_J^2-1)}+C_1\cos\left(\sqrt{\frac{q^2}{q_J^2}-1}\,\tau_*\right)
+C_2\sin\left(\sqrt{\frac{q^2}{q_J^2}-1}\,\tau_*\right).
\ee{k18b}
From the solution (\ref{k18b}) one can infer that for small wavelengths in comparison to Jeans' wavelengths $\lambda_J/\lambda=q/q_J\gg1$ we have time harmonic oscillations  of the density contrast, while  for large wavelengths in comparison to Jeans' wavelengths $\lambda_J/\lambda=q/q_J\ll1$ the density contrast will grow or decay in time. The one which grows is related to Jeans instability.

For the inelastic case where $\alpha\neq1$ we have the following possibilities:
\begin{itemize}
\item For large Jeans' wavelength in comparison with the mean free path $\lambda_J\gg l_0$, eq. (\ref{k17b}) implies that
\ben\lb{k19a}
A_*=\frac43\widetilde T\delta_\rho.
\een
In this case (\ref{k17a}) can be written in terms of the dimensionless time $\tau=t/t_0$ as
\ben\lb{k19b}
\frac{d^2\delta_\rho}{d\tau^2}+\frac79\widetilde T(ql_0)^2\delta_\rho-\frac53(2\pi)^2\frac{l_0^2}{\lambda_J^2}\delta_\rho=0.
\een
The last term of the above equation can be neglected, since it was supposed that $\lambda_J\gg l_0$. From the resulting equation one can infer that for large wavelengths $\lambda$ with respect to the mean free path $ql_0\ll1$ the density contrast has a linear growth with the dimensionless time $\tau$ implying a clustering of the fluid particles due to the inelastic collisions.
\item For the case where Jeans' wavelength is smaller than the mean free path $\lambda_J\ll l_0$, eq. (\ref{k17b}) reduces to $dA_*/d\tau_*=0$ and the equation for the dimensionless temperature  (\ref{k16}) in terms of the dimensionless time $\tau_*$ can be approximate by $\widetilde T(\tau_*)\approx 1$. Hence (\ref{k17a}) reduces to (\ref{k18a}) whose solution is (\ref{k18b}) and we have the same conclusions as above: small wavelengths in comparison to Jeans' wavelengths $\lambda_J/\lambda=q/q_J\gg1$ time harmonic oscillations  of the density contrast occur, while large wavelengths in comparison to Jeans' wavelengths $\lambda_J/\lambda=q/q_J\ll1$ imply that the density contrast will grow or decay in time.  Jeans' instability is related with the growth of the density contrast.
 \end{itemize}

To sum up: (i) instabilities due to the inelastic collisions occur when the Jeans and the perturbation  wavelengths are larger than the mean free path (ii) Jeans' instabilities due to the gravitational field happen when  the mean free path and the perturbation wavelengths are larger than Jeans' wavelength.

For the inelastic case where $\alpha\neq1$ the system of coupled differential equations (\ref{k17a}) and (\ref{k17b})  can be solved numerically by specifying initial conditions for the density  contrast $\delta_\rho$, $A_*$ and fixed values of $\alpha$, $l_0/\lambda_J$ and $q/q_J$. In the numerical simulations which are plotted in the figures below we adopted the initial conditions (say) $\delta_\rho(0)=A_*(0)=0.1$ and ${d \delta_\rho}/{d\tau}(0)=0$ and a fixed value of the normal restitution coefficient $\alpha=0.75$.

\begin{figure}[ht]
\centering
\includegraphics[width=1\textwidth]{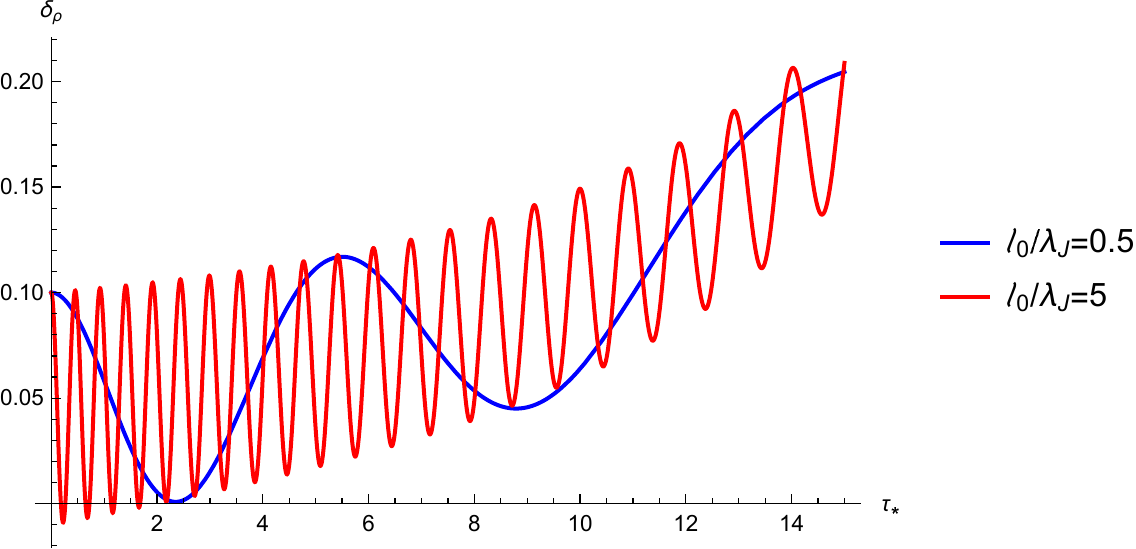}
\medskip
\caption{Density contrast $\delta_\rho$ as function of the dimensionless time $\tau_*$. Small wavelength in comparison to Jeans' wavelength $\lambda_J/\lambda=q/q_J=3$ for the mean free path to Jeans' wavelength ratios $l_0/\lambda_J=0.5$ (blue curve) and $l_0/\lambda_J=5$ (red curve).   }
\label{fig1a}
\end{figure}

\begin{figure}[ht]
\centering
\includegraphics[width=1\textwidth]{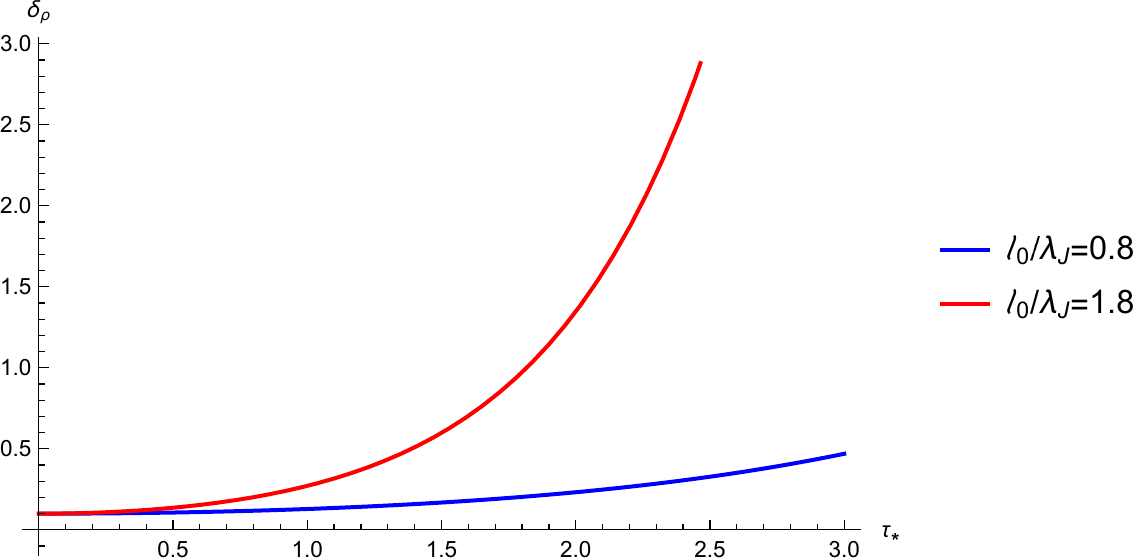}
\medskip
\caption{Density contrast $\delta_\rho$ as function of the dimensionless time $\tau_*$. Large wavelength in comparison to Jeans' wavelength $\lambda_J/\lambda=q/q_J=0.5$ for the  mean free path to Jeans' wavelength ratios $l_0/\lambda_J=0.8$ (blue curve) and $l_0/\lambda_J=1.8$ (red curve).  }
\label{fig1b}
\end{figure}

In Figs. \ref{fig1a} and \ref{fig1b} the density contrast $\delta_\rho$ is plotted as function of the dimensionless time $\tau_*$ by considering the cases of  small and large wavelengths in comparison to Jeans' wavelength, respectively. Fig. \ref{fig1a} represents the small wavelength case where $\lambda_J/\lambda=q/q_J=3$ (say). Here two values of the ratio between the mean free path and Jeans' wavelength $l_0/\lambda_J$ equal to 0.5 and 5  are represented by the blue and red curves, respectively. We can infer from the left frame that the density  contrast execute time  oscillations  which grow with time. By comparing the two curves we note that the  increase of the ratio mean free path to Jeans' wavelength implies a more accentuated oscillation of the density contrast. Note that due to the energy dissipation of the granular gas the period of the oscillations increases.

Large  wavelengths in comparison to Jeans' wavelengths correspond to Jeans' instability where the density contrast grows with time and is shown in  Fig \ref{fig1b} for $\lambda_J/\lambda=q/q_J=0.5$ (say). The blue and red curves refer to the values of the ratio between the mean free path and Jeans' wavelength $\l_0/\lambda_J$ equal to 0.8 and 1.8, respectively.  We note that  by increasing the ratio of the mean free path to Jeans' wavelength a more accentuated time increase of the density contrast occurs.

\section{Hydrodynamic description}

The hydrodynamic  description of a self-gravitating granular gas of smooth inelastic particles is based on the balance equations of mass density $\rho$, hydrodynamic velocity $u_i$ and temperature $T$, namely (see e.g. \cite{Br,Gar2,Kr})
\ben\lb{1a}
&&\frac{\partial\rho}{\partial t}+\frac{\partial\rho u_i}{\partial x_i}=0,
\\\lb{1b}
&&\frac{\partial u_i}{\partial t}+u_j\frac{\partial u_i}{\partial x_j}+\frac1\rho\frac{\partial p_{ij}}{\partial x_j}+\frac{\partial \phi}{\partial x_i}=0,
\\\lb{1c}
&&\frac{\partial T}{\partial t}+u_i\frac{\partial T}{\partial x_i}+\frac{2m}{3k_B\rho}\left[\frac{\partial q_i}{\partial x_i}+p_{ij}\frac{\partial u_i}{\partial x_j}\right]+T\zeta=0,
\een
which are coupled with the Poisson equation (\ref{k6}).
Above $p_{ij}$ is the pressure tensor, $q_i$ the heat flux vector and $\zeta$ the cooling rate.

Equations (\ref{1a}) -- (\ref{1c}) and (\ref{k6}) become a system of field equations for the mass density $\rho$, hydrodynamic velocity $u_i$  and temperature $T$ once the constitutive equations for the pressure tensor $p_{ij}$, heat flux vector $q_i$ and cooling rate $\zeta$ are specified. Here we are interested in analyzing a non-viscous and non-heat conducting Eulerian gas where the pressure tensor reduces to a pressure $p_{ij}=(\rho k_BT/m)\delta_{ij}$ and the heat flux vector vanishes $q_i=0$. For the cooling rate $\zeta$ we shall adopt the expression (\ref{k9b}) derived in the previous section.

For the analysis of the granular gas instabilities we note that the system (\ref{1a}) -- (\ref{1c}) has a background solution corresponding to a constant mass density $\rho_0$, vanishing velocity $u_i^0=0 $, vanishing gravitational potential gradient  $\nabla \phi_0=0$ and a time dependent temperature that obeys Haff's law
${dT}/{dt}+T\zeta=0$. Again we note that the condition of vanishing potential gradient may follow from symmetry properties due to the fact that there is no preferential direction of the gradient in a homogeneous system. As was pointed in the previous section the Poisson equation is not verified with this condition, so that here we shall use the "Jeans swindle" by considering that the Poisson equation is valid only for the perturbed values of the fields.

The solution of Haff's law for the dimensionless temperature $\widetilde T=T/T_0$ as a function of the  dimensionless time $\tau=t/t_0$ is given by (\ref{k16}).
Superposed to the background solution characterized by the fields $\rho_0$, $u_i^0=0$ and $\widetilde T(\tau)$ we add perturbations of small time-dependent amplitudes and space Fourier modes of wavenumber $\bf q$. Here we are interested in analyzing only the longitudinal part of the modes propagating in the $x$-direction so that we write
\ben\lb{4a}
&&\frac{\rho}{\rho_0}=1+\delta_\rho(t)e^{iqx},\qquad u_x=\overline u(t)e^{iqx},
\\\lb{4b}
&& \frac{T}{T_0}=\widetilde T(t)+\delta_T(t)e^{iqx},\qquad \phi=\phi_0+\overline\phi(t)e^{iqx},
\een
where $\delta_\rho(t)$, $\overline u(t)$, $\delta_T(t)$ and $\overline\phi(t)$ are the  time-dependent amplitudes which are considered to be small. The amplitudes $\delta_\rho(t)$ and $\delta_T(t)$ are also known as the density and temperature contrasts, respectively.

Insertion of (\ref{4a}) and (\ref{4b}) into (\ref{1a}) -- (\ref{1c}) and  in the perturbed Poisson equation that follows from (\ref{k6}), leads to the linearized system of equations for the amplitudes
\ben\lb{5a}
\frac{d\delta_\rho}{d t}+iq\overline u=0,
\\\lb{5b}
\frac{d\overline u}{d t}+\frac{k_B}mT_0iq\left(\delta_T+\widetilde T\delta_\rho\right)+iq\overline\phi=0,
\\\lb{5c}
\frac{d\delta_T}{d t}+\frac{2}{3}\widetilde T iq\overline u+\frac{4\rho_0}{3m}\sqrt{\frac{\pi k_BT_0}{m}}\d^2(1-\alpha^2)\widetilde T^\frac12
\left(\frac32\delta_T+\widetilde T\delta_\rho\right)=0,
\\\lb{5d}
-q^2\overline\phi=4\pi G\rho_0\delta_\rho.
\een

From (\ref{5a}) -- (\ref{5d}) we can obtain a coupled system of ordinary differential equations for the   density $\delta_\rho$ and temperature $\delta_T$ contrasts. Indeed,  if we differentiate (\ref{5a}) with respect to time and eliminate the  amplitudes $\overline u$ and $\overline \phi$ by using (\ref{5b}) and (\ref{5d}), respectively, we get
\ben\lb{6a}
\frac{d^2 \delta_\rho}{dt^2}+\frac{k_B}mT_0\left(\delta_T+\widetilde T\delta_\rho\right)q^2-4\pi G\rho_0\delta_\rho=0.
\een
Furthermore the elimination of $\overline u$ from (\ref{5c}) by the use of (\ref{5a}) leads to
\ben\lb{6b}
\frac{d\delta_T}{dt}-\frac23\widetilde T\frac{d\delta_\rho}{dt}+\frac{4\rho_0}{3m}\sqrt{\frac{\pi k_BT_0}{m}}\d^2(1-\alpha^2)\widetilde T^\frac12
\left(\frac32\delta_T+\widetilde T\delta_\rho\right)=0.
\een

As in the previous section, we obtain dimensionless equations from (\ref{6a}) and (\ref{6b}) by introducing the dimensionless time $\tau_*$, the particle mean free path $l_0=m/4\sqrt\pi\rho_0 d^2$ and Jeans's wavenumber $q_J=\sqrt{4\pi G\rho_0}/v_s$ and get
\ben\lb{7a}
&&\frac{d^2 \delta_\rho}{d\tau_*^2}+\left[\frac35\left(\delta_T+\widetilde T\delta_\rho\right)\frac{q^2}{q_J^2}-\delta_\rho\right]=0,
\\\lb{7b}
&&\frac{d\delta_T}{d\tau_*}-\frac23\widetilde T\frac{d\delta_\rho}{d\tau_*}+\frac{1-\alpha^2}{2\pi\sqrt{15}}\frac{\lambda_J}{l_0}\widetilde T^\frac12\left(\frac32\delta_T+\widetilde T\delta_\rho\right)=0.
\een

In the case of elastic collisions  $\alpha=1$, and we get from (\ref{k16}) that  $\widetilde T(0)=1$  and from (\ref{7b}) that $\delta_T=2(\delta_\rho-C)/3$, where $C$ is an integration constant. Hence eq. (\ref{7a}) reduces to (\ref{k18a}) whose solution is (\ref{k18b}).

The case $\alpha\neq1$ has a similar analysis as the one which was done in the previous section. For large Jeans' wavelength in comparison with the mean free path $\lambda_J\gg l_0$ we get from (\ref{7b}) that $\delta_T=-2\widetilde T\delta_\rho/3$ so that eq. (\ref{7a}) in terms of the dimensionless time $\tau$ reduces to
\ben\lb{8a}
\frac{d^2 \delta_\rho}{d\tau^2}+\frac13(ql_0)^2\widetilde T\delta_\rho-\frac{5(2\pi)^2}3\frac{l_0^2}{\lambda_J^2}\delta_\rho\approx \frac{d^2 \delta_\rho}{d\tau^2}+\frac13(ql_0)^2\widetilde T\delta_\rho=0.
\een
From the above equation we get that for large wavelengths $\lambda$ with respect to the mean free path $ql_0\ll1$ a clustering of the fluid particles happens since the density contrast has a linear growth with the dimensionless time $\tau$.

In the case of small Jeans' wavelength in comparison with the mean free path $\lambda_J\gg l_0$ we get from (\ref{7b}) that $3{d\delta_T}/{d\tau_*}=2\widetilde T{d\delta_\rho}/{d\tau_*}$ and it follows the solution (\ref{k18b}).

Let us analyze the numerical solutions of the system of coupled differential equations (\ref{7a}) and (\ref{7b}) for the inelastic case where $\alpha\neq1$. Here we have to specify the initial conditions for the density and temperature contrasts and fixed values of $\alpha$, $l_0/\lambda_J$ and $q/q_J$. In the numerical simulations which are plotted in the figures below we adopted the initial conditions (say) $\delta_\rho(0)=0.5$, $\delta_T(0)=0.1$ and ${d \delta_\rho}/{d\tau}(0)=0$ and fixed values of the normal restitution coefficient $\alpha=0.75$ and of the  mean free path to Jeans' wavelength ratio $l_0/\lambda_J=5$.

\begin{figure}[ht]
\centering
\includegraphics[width=1\textwidth]{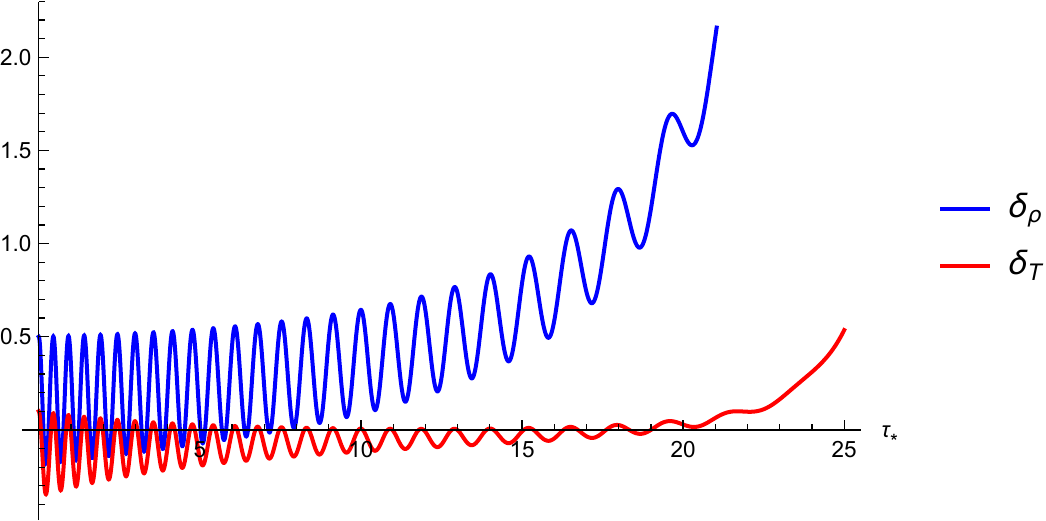}
\medskip
\caption{Density $\delta_\rho$ (blue) and temperature $\delta_T$ (red) contrasts as functions of the dimensionless time $\tau_*$ for fixed value of the mean free path to Jeans' wavelength ratio $l_0/\lambda_J=5$. Small wavelength in comparison to Jeans' wavelength $\lambda_J/\lambda=q/q_J=3$. }
\label{fig2a}
\end{figure}

\begin{figure}[ht]
\centering
\includegraphics[width=1\textwidth]{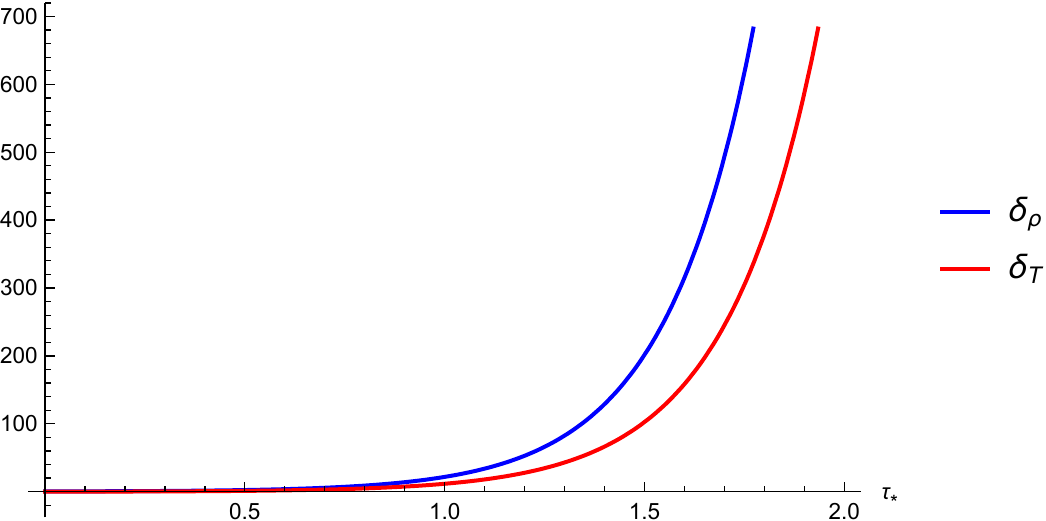}
\medskip
\caption{Density $\delta_\rho$ (blue) and temperature $\delta_T$ (red) contrasts as functions of the dimensionless time $\tau_*$ for fixed value of the mean free path to Jeans' wavelength ratio $l_0/\lambda_J=5$. Large wavelength in comparison to Jeans' wavelength $\lambda_J/\lambda=q/q_J=0.5$. }
\label{fig2b}
\end{figure}

The density $\delta_\rho$ and temperature $\delta_T$ contrasts are plotted in Figs. \ref{fig2a} and \ref{fig2b} as functions of the dimensionless time $\tau_*$. In these figures the blue curves represent the density contrast and the red one the temperature contrast. Figure \ref{fig2a}  refers to the case of small wavelengths in comparison to Jeans' wavelengths $\lambda_J/\lambda=q/q_J=5$  and we observe that the density and temperature contrasts execute time  oscillations which grow with time, moreover the temperature contrast has a less accentuated time evolution. The case of large wavelengths in comparison to Jeans' wavelengths correspond to Jeans' instability and it is shown in Fig. \ref{fig2b} for $\lambda_J/\lambda=q/q_J=0.5$. We note that  the density contrast has a more accentuated time increase with respect to  the temperature contrast.

\section{Final Remarks}

As molecular clouds are the sites of star formation, their formation,
internal structure and dynamics determines the rate of star formation
and the properties of young stars. For that reason, we study the
emergence of the Jeans instability in a molecular gas  with some
cooling effect  encoded in the Haff's law.  To do so, we begin with
some basic facts.  Let us consider a region with radius $R$ and
enclosed mass $M$ with some baryonic components and mean mass density
$\rho$.  To take into account the cooling effect in  the gas, we must
consider that the gas can radiate energy  and cool in such a way that
the matter will collapse further  and form a bound object. When
studying Jeans instability with cooling effect, we basically have
three different kinds of time-scale. We have the dynamical time scale
or free-fall time defined as
$t_{dyn}= (2GM/R^{3})^{-1/2} $,
the  time-scale defined with the pressure as
$t_{press}= \lambda/v_{s} $,
where the sound speed is $v_{s}=\sqrt{p/\rho}$. The cooling time scale due to the Haff's law $\dot{T}=-\zeta T$,  
can be taken as an approximation as
$t_{cool}=\zeta^{-1}$. Note that  the cooling rate depends essentially of the
temperature and  density (\ref{k9b})$_2$. 
In order to get Jeans instability, we must demand that
$t_{dyn}>t_{press}$ or equivalently
$\lambda>\lambda_{J}=v_{s}/\sqrt{G\rho}$. After the instability
appeared the next question is what happens with that structure, it will 
further break in smaller pieces or not (fragmentation process). In
order to give an answer  we must see what is the role of the  cooling
time in the evolution of the molecular cloud gas. To do, it is useful
to distinguish three different possibilities.  If $t_{cool}$ is much
bigger than the Hubble time scale $t_{H}=H^{-1}$ the cloud could not
evolve much since its formation, so we discard this possibility. If
$t_{H}>t_{cool}>t_{dyn}$ the  gas can cool  but keeping the pressure
adjustable in order to have a cloud quasi static on a time scale of
order $t_{cool}$. If $t_{cool}<t_{dyn}$ the molecular cloud will cool
until reaches its  minimum temperature, the loss of pressure will lead
to a free-fall collapse type. In this way, the fragmentation process
can proceed to smaller mass scales. The criterion $t_{cool}<t_{dyn}$
is useful to determine the masses of galaxies provided only when this
condition is met,  the gravitating molecular cloud can  collapse  and
fragment, creating a star. To see how important the cooling effects
can be we can estimate the luminosity emitted by the configurations.
The luminosity is the energy emitted over  some time, say $L= E t^{-1}$=
[area of the detector]$\times$[flux of the source]. In this case the
average energy of molecular configuration is $E_{dyn} \simeq G
M^{2}/R$  and the free-fall time scale is $t_{dyn}=
R^{3/2}(2GM)^{-1/2}$ then  the average luminosity associated with
this time-scale is $L_{dyn}=E_{dyn}t^{-1}_{dyn}$. In the same way, the
average luminosity associated with the cooling time-scale is
$L_{cool}=E_{cool}t^{-1}_{cool}$. In order to compare both stages, let
us assume that the average energy is reduced due to the cooling
process, that is, $E_{cool}=\mathcal{P}E_{dyn}$ with $0<\mathcal{P}<1$. Then, the
criterion $t_{cool}<t_{dyn}$ implies that the consistency relation
$\mathcal{P}<L_{cool}/L_{dyn}<1$. Notice that this condition can be verified if
it is correct provided we have $t_{dyn}$ and $t_{cool}$. Using that
$\rho=M/R^{3}$, $k_BT=GM^{2}/R$, we arrived at
$t_{cool}=\zeta^{-1}\propto \rho\sqrt{k_BT}=R^{7/2}/G^{1/2}M^2$  and the criterion $t_{cool}<t_{dyn}$ leads
to a relation between mass and radius given by $\beta
R^2<M^{3/2}$ where $\beta$ involves some constants, but it is
fixed. In this way we can look the range of $R$ and $M$ that are
consistent with the well established theoretical bounds, see for
instance  Padmanabhan \cite{Pad}.  There,
it was shown that for a more realistic model with another  cooling
rate  for temperature $< 10^{6}$K  the relation  $t_{cool}<t_{dyn}$
implies  the efficient cooling is achieved as long as
$M<10^{12}M_{\odot}$. For $T> 10^{6}$K clouds can form galaxies
only if they shrink below the $10^{2}$kpc.

\section{Conclusions}

In this work we have investigated a self-gravitating granular gas  in order to analyze the    clustering of the fluid particles due to the  instabilities generated by the inelastic collisions and by the gravitational field. Two methodologies were used: in the first one the Boltzmann equation was coupled with the Poisson equation, while in the second one the Poisson equation was coupled with the macroscopic balance equations of mass density, hydrodynamic velocity and temperature for an Eulerian fluid, where the only  dissipative effect was the temperature decay due to the inelastic collisions of the fluid particles. For both cases  the background solution  was characterized by a fluid  at rest with constant mass density, time dependent temperature and constant gravitational potential. The time dependent temperature was given by Haff's law of the homogeneous cooling solution.

In the kinetic description the perturbed distribution function was represented as plane waves of fixed wavenumber and time dependent small amplitudes associated with the mass, momentum and energy of the fluid particles. The perturbed gravitational potential was a function of the wavenumber and a time dependent small amplitude. A coupled system of  differential equations for the amplitudes was obtained from the Boltzmann and  Poisson equations and the main objective was the determination of  the time evolution of the density contrast, which is a parameter that indicate where there are local enhancements in the matter density.    It was shown that instabilities due to the inelastic collisions occur when the Jeans and the perturbation  wavelengths are larger than the mean free path, while Jeans' instabilities due to the gravitational field happen when  the mean free path and the perturbation wavelength are larger than Jeans' wavelength.

In the hydrodynamic description the fields of mass density, hydrodynamic velocity and temperature were perturbed from the background solution by considering the perturbations in the Fourier space with time dependent amplitudes.
From the balance and Poisson equations a coupled system of differential equation for the amplitudes was also obtained. The analysis of the instabilities was based on a coupled system of differential equations for the density and temperature contrasts. As was expected the same conclusions were obtained as those of the kinetic description for the density contrast. Here it was shown that the time evolution of the temperature contrast evolve more slowly than the density contrast.

\section*{Acknowledgments} The support of Conselho Nacional de Desenvolvimento Cient\'ifico e Tecnol\'ogico  (CNPq -- Brazil). I would like to thank Professors Andr\'es Santos and Vicente Garz\'o to called my attention to the paper \cite{b11} and Dr. Mart\'in Richarte for his help in the astrophysical topic.

\end{document}